# Little-Parks oscillations with half-quantum fluxoid features in $Sr_2RuO_4$ micro rings

Yuuki Yasui,[1,*] Kaveh Lahabi,[2] Muhammad Shahbaz Anwar,[1,3] Yuji Nakamura,[1]
Shingo Yonezawa,[1] Takahito Terashima,[1] Jan Aarts,[2] and Yoshiteru Maeno[1]

[1]*Department of Physics, Graduate School of Science, Kyoto University, Kyoto 606-8502, Japan*
[2]*Huygens-Kamerlingh Onnes Laboratory, Leiden University, P.O. Box 9504, 2300 RA Leiden, The Netherlands*
[3]*London Centre for Nanotechnology, University College London, 17-19 Gordon Street, London, WC1H 0AH, United Kingdom*

In a micro ring of a superconductor with a spin-triplet equal-spin pairing state, a fluxoid, a combined object of magnetic flux and circulating supercurrent, can penetrate as half-integer multiples of the flux quantum. A candidate material to investigate such half-quantum fluxoids is $Sr_2RuO_4$. We fabricated $Sr_2RuO_4$ micro rings using single crystals and measured their resistance behavior under magnetic fields controlled with a three-axes vector magnet. Proper Little-Parks oscillations in the magnetovoltage as a function of an axially applied field, associated with fluxoid quantization are clearly observed, for the first time using bulk single crystalline superconductors. We then performed magnetovoltage measurements with additional in-plane magnetic fields. By carefully analyzing both the voltages $V_+$ ($V_-$) measured at positive (negative) current, we find that, above an in-plane threshold field of about 10 mT, the magnetovoltage maxima convert to minima. We interpret this behavior as the peak splitting expected for the half-quantum fluxoid states.

Recently, it has been recognized that Majorana particles, which have unusual equivalence to their own antiparticles and have been long sought in elementary particle physics, can be realized as quasiparticle excitation in condensed-matter systems such as topological superconductors [1]. In particular, Majorana zero modes (MZMs), the zero-energy states of the Majorana branch, have attracted much attention since MZMs do not obey ordinary Abelian statistics and can be utilized for quantum computing [2, 3]. Thus, direct detection of MZMs has become a holy grail of current condensed matter physics [4, 5]. Half-quantum fluxoid (HQF) [6] in a spin-triplet superconductor or a superfluid is known to host such MZMs [7, 8].

An additional phase degree of freedom in a superconducting wave function is the key ingredient for the realization of HQF states. For a spin-singlet superconducting ring with wave function $\psi_S = |\Delta_S|e^{i\theta}$, the single-valuedness of $\psi_S$ requires quantization $\Phi' = n\Phi_0$ (integer-quantum fluxoid, IQF) inside a closed path. Here, $n$ is an integer, $\Phi'$ is the fluxoid, and $\Phi_0 = h/2e$ is the flux quantum with $h$ the Planck constant and $e$ the elementary charge. Note that, in a superconductor smaller than the penetration depth, the fluxoid, which contains an integration of the supercurrent along a closed path, is quantized, rather than the flux. For a spin-triplet equal-spin pairing (ESP) superconductor, the wavefunction $\psi_T = |\Delta_T|(-e^{i\theta_\uparrow}|\uparrow\uparrow\rangle + e^{i\theta_\downarrow}|\downarrow\downarrow\rangle)$ has two phase degrees of freedom. In an ESP ring, half-integer quantization $\Phi' = n'\Phi_0$ with $n' = \pm 1/2, \pm 3/2, \cdots$ is allowed even under the constraint of the single-valuedness of the wave function. Such a fluxoid state is called the HQF state.

One of the materials that can host the HQF is $Sr_2RuO_4$, which is a leading candidate spin-triplet ESP superconductor [9, 10]. This oxide has a layered perovskite structure and exhibits superconductivity below 1.5 K. Various experiments

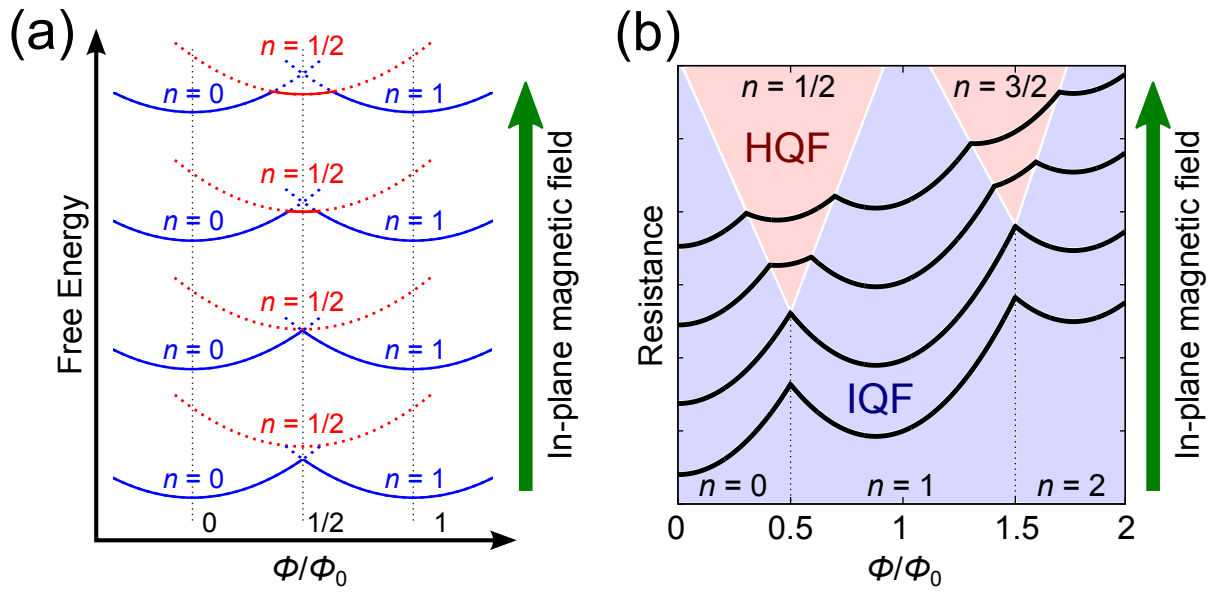

FIG. 1. (a) Schematic profile of the free energies for IQF states and a HQF state. A HQF state may become energetically favorable under in-plane magnetic field; it is realized above a threshold in-plane field value. (b) Expected change of the magnetoresistance oscillations. Peak splittings with in-plane field are expected when HQF states are realized.

have provided firm evidence for the spin-triplet ESP state [11–18], but there still are several issues that cannot be understood within the current spin-triplet scenario [19–23]. Also, signatures of HQF have been observed in microstructured $Sr_2RuO_4$ rings using cantilever magnetometry [24]. Still, in order to come to Majorana braiding, electrical detection of the HQF state is required. For this, small samples are necessary in order to reduce the spin-current energy of the HQF state, as pointed out by Chung *et al.* [25]. The role of the Zeeman field to lower the kinetic energy of a HQF state is discussed by Vakaryuk and Leggett [26]. Accordingly, a transition from IQF to HQF is expected to occur with increasing in-plane field, with free-energy minima for the HQF states appearing in the middle of neighboring IQF states [Fig. 1(a)]. There is a proposal for detection of HQF using perforated films [27]; here we use a simpler system of a ring shape. It should also be mentioned that a superconducting state can exist with an enhanced transition temperature, the so-called 3-K phase. This is observed in eutectic crystals [28–30] or in bulk crystals under uniaxial

* y-yasui@scphys.kyoto-u.ac.jp

strain [31–33].

Fluxoid quantization can be investigated by measuring magnetoresistance oscillations as a function of a field applied along the axis of the ring in the regime of the resistive transition, known as the Little-Parks (LP) oscillations [34]. The LP oscillations originate from the oscillations in the free energy and hence in the transition temperature $T_c$, caused by field-induced supercurrents that flow to satisfy the quantization condition. Then the magnetoresistance curve should trace the field dependence of the free energy [Fig. 1]. Thus, a resistance peak in the LP oscillations, located at the border of two neighboring IQF states, should split when HQF states are realized, as shown in Fig. 1(b). There are indications that the order parameter in the 3-K phase is not ESP [33, 35]. However, the LP oscillations are robust, regardless of the pairing symmetry or the number of components of the order parameter.

Although techniques to detect the LP oscillations have been developed over the past 50 years, all reported experiments have been performed using superconducting films. To the best of our knowledge, there is no report of the observation of proper LP oscillations even for IQF in a ring made of bulk single crystals, including $Sr_2RuO_4$. For $Sr_2RuO_4$, although its superconducting thin films have been reported [36, 37], films with strong and homogeneous superconductivity are still virtually absent. Therefore, for LP experiments, techniques to make micro rings without using thin films are needed. Recently, Cai *et al.* reported observation of magnetoresistance oscillations in micro rings made of single-crystalline $Sr_2RuO_4$ [38, 39]. However, the oscillation amplitudes were substantially larger than the LP expectation.

Here, we report the first observation of proper Little-Parks oscillations in micro rings of single-crystalline $Sr_2RuO_4$. With in-plane fields, we observed two different kinds of peak splittings of the LP oscillations: after careful examination of the raw voltage, we conclude that the splitting in small in-plane fields is extrinsic, originating from asymmetry in the current-voltage characteristics; whereas the splitting in larger in-plane fields, observable also in the raw voltage, is intrinsic.

In this study, $Sr_2RuO_4$ single crystals grown with the floating-zone method [40] were used for micro rings. Before the fabrication, $T_c$ of the crystal C391, used for Sample B, was confirmed to be 1.50 K [Fig. S4] using AC susceptibility method (Quantum Design, PPMS adiabatic demagnetization refrigerator option) [41]. A 1-$\mu$m-thin crystal was selected among crushed single crystals, and it was placed on a $SrTiO_3$ substrate, which has a thermal contraction matching with that of $Sr_2RuO_4$. (For Sample A, however, a sapphire substrate with a smaller thermal contraction was used.) The surface of the crystal was protected by evaporating a thin layer of $SiO_2$ after electrodes of high-temperature-cure silver paint (Dupont, 6836) are provided. To fabricate rings with a four-terminal configuration [Figs. 2(a)-(c)], the Ga-based focused ion beam (FIB) technique was used with a 20-pA and 30-kV beam. The rings were cooled down to 0.3 K with a $^3$He refrigerator (Oxford Instruments, Heliox). To avoid influence of thermoelectric voltage, the resistance was measured under DC current with sign flip: We measure voltage under positive current $V_+ = V(+I)$ and under negative current $V_- = V(-I)$, and evaluate the resistance $R$ as $R = (V_+ - V_-)/2I$. To investigate the field dependence, however, it is crucially important to examine $V_+$ and $V_-$ individually since LP oscillations are not necessarily invariant under reflection, as seen below. In other words, it is essential to examine magnetovoltage rather than magnetoresistance. Temperature stability during a magnetotransport measurement is approximately 100 $\mu$K. This value is substantially smaller than the expected transition-temperature shift due to the LP oscillations, estimated to be around 10 mK. The magnetic field was applied with a three-axes superconducting vector magnet (1 T / 0.2 T / 0.2 T), allowing us to control the out-of-plane and in-plane fields independently. More details on the experimental method are described on the Supplemental Material [42].

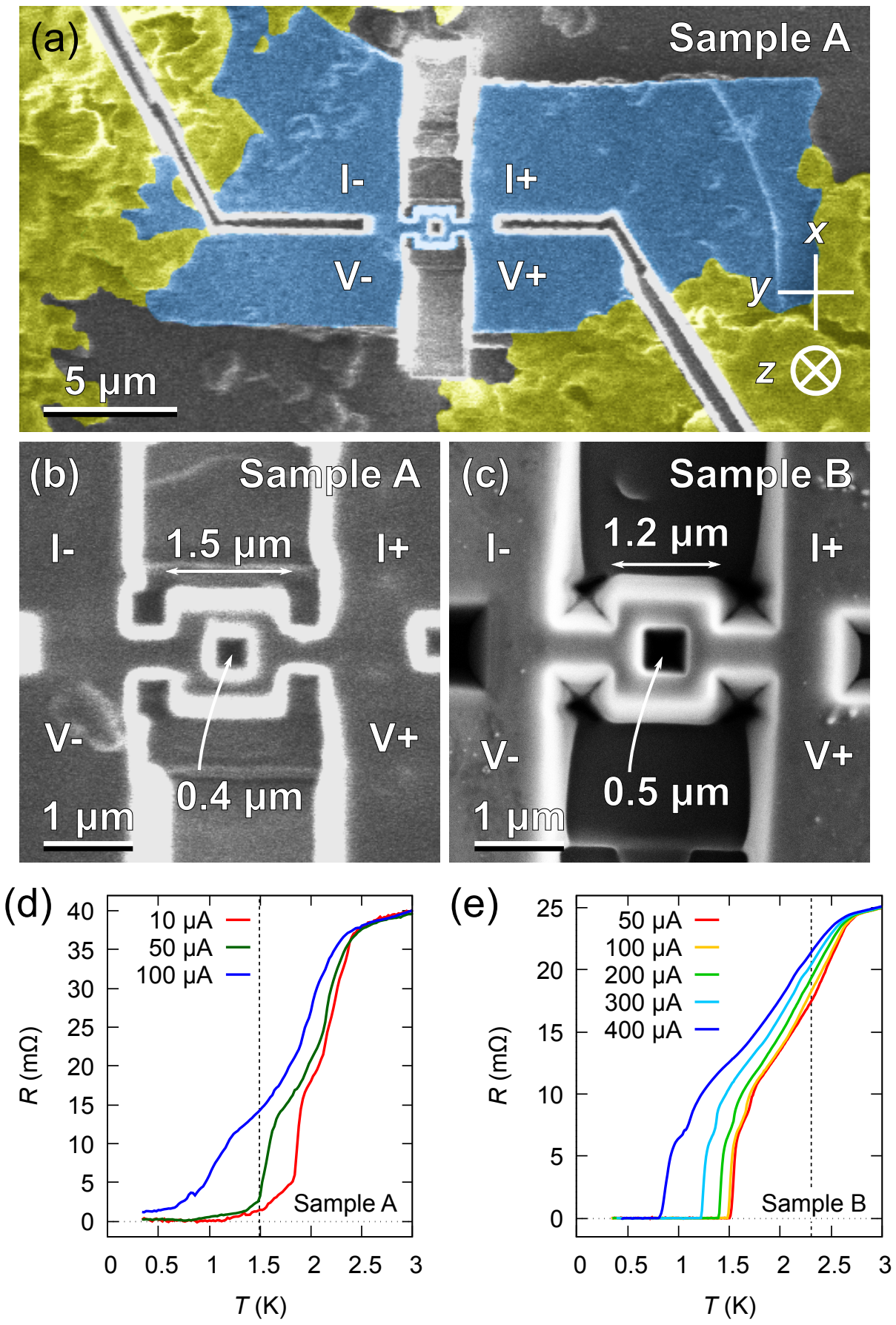

FIG. 2. (a) False-colored scanning ion microscope (SIM) image of Sample A (yy075). The blue- and yellow-colored regions are the $Sr_2RuO_4$ crystal and the silver paint, respectively. (b) Magnified SIM image of Sample A. (c) Scanning electron microscope (SEM) image of Sample B (yy150). The thickness of Samples A and B are 1.3 and 2.0 $\mu$m, respectively. Resistance of (d) Sample A and (e) Sample B as functions of temperature. Both rings show superconducting transition above 1.5 K and several other transition steps. The dashed lines indicate the temperatures at which the magnetoresistance and magnetovoltage shown in Fig. 3 are measured.

Figures 2(d) and (e) show the temperature dependence of the ring resistance $R(T)$. Zero resistance due to superconductivity was observed in both rings. Note that the superconducting transitions start well above 1.5 K. This is a signature of the 3-K phase, which is induced in our rings probably by local strain, caused by the FIB process. Several transition steps are observed in both rings. Each step corresponds to the transition of a certain region of the device, as demonstrated in a LP experiment using a conventional superconductor [43]. Still, the correspondence is not entirely straightforward and we rather identify the contribution from the ring by finding the temperature regime where field-induced resistance oscillations occur. In Sample A this is around 1.5 K. In Sample B it is around 2.5 K, while the transitions below 2 K probably are connected to the neck part and the contact part of the structure [Fig. S6].

The magnetoresistance of the rings in the regions of the resistive transitions is shown in Figs. 3(a) and (b). The measurements were performed at fixed temperatures indicated with dashed lines in Figs. 2(d) and (e). The samples were heated above 5 K before each measurement, then cooled under zero magnetic field. Periodic oscillations were observed with periods $\mu_0\Delta H = 2.6$ mT and 3.8 mT for Samples A and B, respectively. From $\Delta H$, we can estimate the area $S$ that causes the oscillations by using the relation $\Phi_0 = \mu_0\Delta H \cdot S$. As a result, we obtain $S_{\mathrm{SampleA}} = 0.80\,\mu\mathrm{m}^2$ and $S_{\mathrm{SampleB}} = 0.54\,\mu\mathrm{m}^2$, which agree well with the geometry of the rings.

Next, we quantitatively evaluate the oscillations. The shift of $T_\mathrm{c}$ due to the fluxoid quantization is given by [44]

$$\frac{T_\mathrm{c}(H) - T_\mathrm{c}(0)}{T_\mathrm{c}(0)} = -\left(\frac{\pi\xi_0 w\mu_0 H}{\sqrt{3}\Phi_0}\right)^2 - \frac{\xi_0^2}{r_1 r_2}\left(n - \frac{\pi\mu_0 H r_1 r_2}{\Phi_0}\right)^2, \tag{1}$$

where $\xi_0$ is the coherence length at 0 K, $r_1$ is the inner radius, $r_2$ is the outer radius, and $w = r_2 - r_1$ is the width of a ring. In our calculation, $\xi_0 = 66$ nm (the coherence length along the $ab$-plane of $Sr_2RuO_4$ [10]), while we chose $2r_1 = 0.75\,\mu$m, $2r_2 = 1.4\,\mu$m for Sample A, and $2r_1 = 0.7\,\mu$m, $2r_2 = 1.0\,\mu$m for Sample B. Note that the samples are somewhat "conical", with a smaller top and larger bottom. To convert the $T_\mathrm{c}$ shift to a resistance shift, we assume that the shape of a $R(T)$ curve does not change under magnetic field, and the curve shifts to the left by $T_\mathrm{c}(0) - T_\mathrm{c}(H_z)$. As presented in Figs 3(a) and (b) the obtained $R(H_z)$ simulations agree well with the experimental results without any adjustable parameters. We observed oscillations corresponding to $|n| \leq 5$ for Sample A and $|n| \leq 3$ for Sample B. This is because the parabolic component due to the Meissner effect (the first term in Eq. (1)) is dominant at a high field region, and the oscillation component (the second term in Eq. (1)) is not resolved. Though a modulation of the oscillatory period is known in a wide-arm ring [45], we do not observe such non-periodic oscillations. We emphasize again that we succeeded in observing the LP oscillations using a bulk single crystal unlike the other reported LP experiments using superconducting films [46]. Thus, the first conclusion of this paper is that the magnetoresistance oscillations observed in both $Sr_2RuO_4$ micro rings are the proper LP oscillations.

We then performed magnetotransport measurements with additional in-plane magnetic fields $H_y$ (which is along the current direction). The magnetoresistance as well as the raw voltages $V_+$ and $V_-$ for $\mu_0 H_y = 8$ and 20 mT are shown in Fig. 3(c). The out-of-plane magnetic field values were corrected for the misalignment of the rings with respect to the magnets. To be specific, the actual out-of-plane field $H_z$ is given by $H_z = H_z^{\mathrm{magnet}}\cos\theta + H_y^{\mathrm{magnet}}\sin\theta + H_z^{\mathrm{remnant}}$, where the misalignment angle $\theta = 0.86^\circ$ and the remnant field $H_z^{\mathrm{remnant}} = -0.3$ mT are chosen so that the peaks are located at the same $|H_z|$ value [47].

For $\mu_0 H_y = 8$ mT, the magnetoresistance $R(H_z)$ peaks appear split at $\mu_0 H_z = \pm 1.3$ mT, which correnpond to the transition fields between $n = 0$ and $n = \pm 1$ fluxoid states. However, this peak splitting is not observed in the magnetovoltage $V_+(H_z)$ or $V_-(H_z)$. Instead, the peaks for $V_+(H_z)$ and $V_-(H_z)$ emerge at different $H_z$. Notice that, the resistance is obtained from an average of $V_+$ and $-V_-$. As a result, the difference of the peak position in $V_+(H_z)$ and $-V_-(H_z)$ causes artifact peak splitting in the magnetoresistance. Thus, to find an intrinsic peak splitting originating from HQFs, not only $R(H_z)$ but also $V_+(H_z)$ and $V_-(H_z)$ data should be carefully examined: current-averaged resistance data may cause misinterpretation of experimental results.

For $\mu_0 H_y = 20$ mT the situation is different. In this case the splitting in $R(H_z)$ is also observed in $V_+(H_z)$ (see the top two panels of Fig. 3(c)). Thus, this splitting is not an artifact originating from the asymmetric peaks in $V_+(H_z)$ and $V_-(H_z)$. In the rest of this paper, we focus on this splitting in the magnetovoltage.

Figure 3(d) shows the $V_+(H_z)$ with 4-mT $H_y$ steps. Under zero in-plane magnetic field, the oscillations are consistent with the ordinary LP magnetovoltage oscillations with a period corresponding to $\Phi_0$. When the in-plane field is applied above 12 mT, the peaks in $V_+(H_z)$ clearly start to split. Furthermore, the width of the splitting becomes larger with increasing in-plane field. The increased splitting is consistent with the expectation that the free energy of a HQF state becomes smaller under the in-plane field, as shown in Fig. 1. Interestingly, the dips at $\mu_0 H_z = \pm 1.3$ mT for $\mu_0 H_y = 20$ mT are even deeper than the voltage bottoms of the IQF states. Within the HQF scenario, this suggests that the energy of HQF states can become smaller than that of IQF states. We emphasize that the results are well reproducible. The measurements were repeated twice in each condition, and the obtained curves precisely match each other. Magnetoresistance measurements with another in-plane field direction and on Sample B were also performed [42]. For Sample B we do not see signatures of the HQF state in the field range where we expect them, although some sort of splitting occurs above 150 mT.

It may be argued that, if several transition steps in $R(T)$ contribute to the $V_+(H_z)$ and $V_-(H_z)$, the voltages may exhibit a complicated shape resembling that of a HQF state. However, even with 20 mT in-plane field, the resistance is still lower than 6 m$\Omega$ as shown in the upper panel of Fig. 3(c). Figure 2(d) shows that the resistance corresponding to the lowest-temperature transition is $R < 10$ m$\Omega$. Therefore, magnetoresistance measurements were always performed at the sharp transition region occurring around 1.5 K and the higher-temperature transitions do not contribute the magneto-transport.

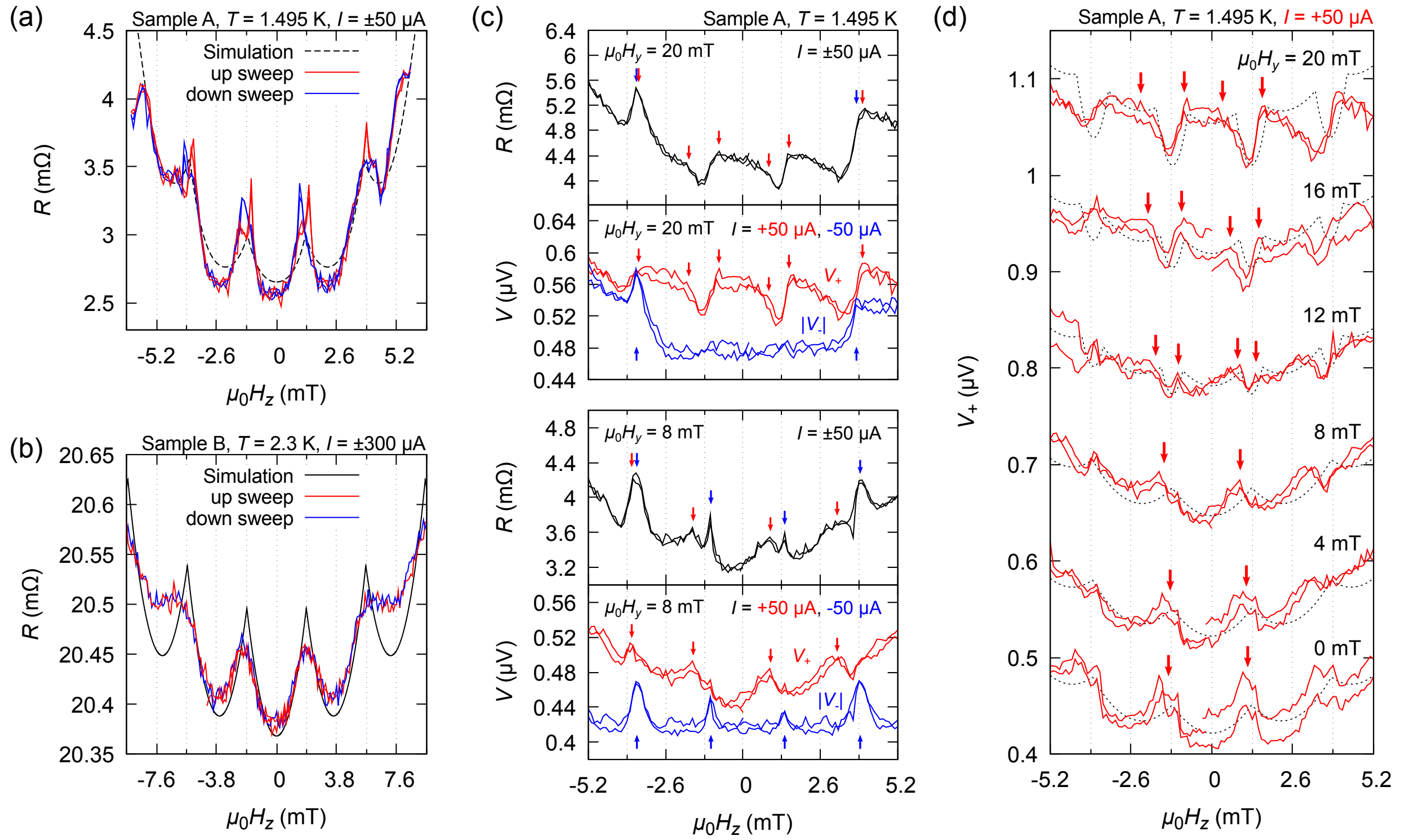

FIG. 3. Magnetoresistance $R(H_z)$ of (a) Sample A and (b) Sample B without in-plane fields. Both oscillatory periods and amplitudes agree with those of simulations for the Little-Parks oscillations. (c) Comparison of resistance and voltage as functions of $H_z$ for up-sweeps at 0.3 mT/min under constant in-plane fields $H_y$. At $\mu_0 H_y$ = 8 mT, the difference in the peak positions in $V_+(H_z)$ and $V_-(H_z)$ results in apparent resistance-peak-splitting because the resistance is evaluated as $(V_+(H_z) - V_-(H_z))/2I$. For $\mu_0 H_y$ = 20 mT, however, dips at $\mu_0 H_z = \pm 1.3$ mT are clearly observed even in the raw voltage $V_+$. Hence the HQF dips in the resistance is not an artifact originating from averaging. Note that for $V_-$, its absolute values $|V_-|$ are plotted with vertical offsets. (d) Effects of in-plane magnetic field $H_y$ on the magnetovoltage $V_+(H_z)$ of Sample A, including the data shown in (c). Magnetovoltage peaks split above $\mu_0 H_y$ = 12 mT as indicated with arrows, and the width of the splitting becomes wider with increasing $H_y$, as expected for HQF states. Measurements are repeated twice in each conditions to demonstrate good reproducibility. Each set of curves has a 0.1-$\mu$V offset for clarity. The dashed curves are guide to the eye.

Let us here compare our results with the previous cantilever magnetometry study by Jang *et al.* [24]. Their measurements were performed at 0.6 K, much lower than $T_c$. In contrast, our experiment was conducted around $T_c$ to measure finite resistance/voltage. Besides, the measurement current may interact with the circulating supercurrent in our measurements. In spite of these differences, additional features at $\pm\Phi_0/2$ are present in both experiments. Moreover, in both cases the HQF features are only observed with $\mu_0 H_y$ above around 10 mT. Together, the data suggest that the HQF states are very likely to be intrinsic to $Sr_2RuO_4$.

There are still issues to be resolved. First, hysteresis is observed in the $H_z$ sweep [Figs. S6-S9]. Such hysteresis between fluxoid states may occur because of the metastable branches in the free energy (dotted parts of the curves in Fig. 1(a)). Nevertheless, a detailed mechanism for the asymmetric hysteresis especially at large $H_y$ is still unclear. We comment here that similar hysteresis was also observed in the torque experiment [24]. Second, the splittings of the magnetovoltage peaks for positive $H_y$ are observed only in $V_+$ but not in $V_-$. Nevertheless, for negative $H_y$, peaks in $V_-$ show splitting but not in $V_+$ [Figs. S6 and S7]. This result ensures the expected symmetry under the concurrent inversion of magnetic field and current: $\boldsymbol{H} \to -\boldsymbol{H}$ and $\boldsymbol{I} \to -\boldsymbol{I}$, $V_\pm(\boldsymbol{H}) \simeq -V_\mp(-\boldsymbol{H})$. On the other hand, the dips in voltages are affected under $y$-direction field inversion: $H_y \to -H_y$, $V_\pm(H_z, H_y) \neq V_\pm(H_z, -H_y)$. Perhaps, one needs to consider the role played by the geometrical asymmetry, for example inhomogeneity in $T_c$ or difference in the effective width between the positive- and negative-$x$ halves of the ring. Finally, the question can be raised why we do not observe the large magnetoresistance oscillations seen by Cai *et al.* [38, 39]. We have also investigated circular rings (rather than the square ones discussed here), and observed no LP oscillations but large amplitude magnetoresistance oscillations. A detailed comparison and its possible origin will be discussed in a subsequent paper.

In conclusion, we have observed the LP oscillations with expected amplitudes and periods in micro rings of $Sr_2RuO_4$. This is the first report of the LP oscillations using any bulk single-crystal superconductor. Furthermore, by applying in-plane magnetic fields, we observed splitting of the peaks of the LP magnetovoltage oscillations. The widening of the splitting with increasing in-plane field agrees with the expectation for the HQF state.

## ACKNOWLEDGMENTS

We would like to acknowledge Y. Yamaoka for his technical contribution, Z. Q. Mao and F. Hübler for their contributions to crystal growth, P. F. A. Alkemade for his help with sample structuring, and R. Budakian for his advice on ring preparation. This work was supported by a Grant-in-Aid for Scientific Research on Innovative Areas "Topological Materials Science" (KAKENHI Grant Nos. JP15H05852, JP15K21717, JP15H05851), JSPS Core-to-Core program, JSPS research fellows (KAKENHI Grant Nos. JP16J10404, 26-04329), and Grant-in-Aid JSPS KAKENHI JP26287078 and JP17H04848. It was also supported by the Netherlands Organisation for Scientific Research (NWO/OCW), as part of the Frontiers of Nanoscience program. YY would like to acknowledge support from the Motizuki Fund of Yukawa Memorial Foundation.

[1] X.-L. Qi and S.-C. Zhang, Rev. Mod. Phys. **83**, 1057 (2011).
[2] D. A. Ivanov, Phys. Rev. Lett. **86**, 268 (2001).
[3] M. Stone and S. B. Chung, Phys. Rev. B **73**, 014505 (2006).
[4] V. Mourik, K. Zuo, S. M. Frolov, S. R. Plissard, E. P. A. M. Bakkers, and L. P. Kouwenhoven, Science **336**, 1003 (2012).
[5] S. Nadj-Perge, I. K. Drozdov, J. Li, H. Chen, S. Jeon, J. Seo, A. H. MacDonald, B. A. Bernevig, and A. Yazdani, Science **346**, 602 (2014).
[6] G. Volovik and V. Mineev, JETP Letters **24**, 561 (1976).
[7] N. Read and D. Green, Phys. Rev. B **61**, 10267 (2000).
[8] C. Kallin and J. Berlinsky, Reports on Progress in Physics **79**, 054502 (2016).
[9] A. P. Mackenzie and Y. Maeno, Rev. Mod. Phys. **75**, 657 (2003).
[10] Y. Maeno, S. Kittaka, T. Nomura, S. Yonezawa, and K. Ishida, J. Phys. Soc. Jpn. **81**, 011009 (2012).
[11] K. Ishida, H. Mukuda, Y. Kitaoka, K. Asayama, Z. Q. Mao, Y. Mori, and Y. Maeno, Nature **396**, 658 (1998).
[12] G. Luke, Y. Fudamoto, K. Kojima, M. Larkin, J. Merrin, B. Nachumi, Y. Uemura, Y. Maeno, Z. Mao, Y. Mori, H. Nakamura, and M. Sigrist, Nature **394**, 558 (1998).
[13] J. A. Duffy, S. M. Hayden, Y. Maeno, Z. Mao, J. Kulda, and G. J. McIntyre, Phys. Rev. Lett. **85**, 5412 (2000).
[14] J. Xia, Y. Maeno, P. T. Beyersdorf, M. M. Fejer, and A. Kapitulnik, Phys. Rev. Lett. **97**, 167002 (2006).
[15] S. Kashiwaya, H. Kashiwaya, H. Kambara, T. Furuta, H. Yaguchi, Y. Tanaka, and Y. Maeno, Phys. Rev. Lett. **107**, 077003 (2011).
[16] M. S. Anwar, S. R. Lee, R. Ishiguro, Y. Sugimoto, Y. Tano, S. J. Kang, Y. J. Shin, S. Yonezawa, D. Manske, H. Takayanagi, T. W. Noh, and Y. Maeno, Nat. Commun. **7**, 13220 (2016).
[17] M. Manago, K. Ishida, Z. Mao, and Y. Maeno, Phys. Rev. B **94**, 180507 (2016).
[18] M. Manago, T. Yamanaka, K. Ishida, Z. Mao, and Y. Maeno, Phys. Rev. B **94**, 144511 (2016).
[19] C. W. Hicks, J. R. Kirtley, T. M. Lippman, N. C. Koshnick, M. E. Huber, Y. Maeno, W. M. Yuhasz, M. B. Maple, and K. A. Moler, Phys. Rev. B **81**, 214501 (2010).
[20] S. Yonezawa, T. Kajikawa, and Y. Maeno, Phys. Rev. Lett. **110**, 077003 (2013).
[21] S. Yonezawa, T. Kajikawa, and Y. Maeno, J. Phys. Soc. Jpn. **83**, 083706 (2014).
[22] S. Kittaka, A. Kasahara, T. Sakakibara, D. Shibata, S. Yonezawa, Y. Maeno, K. Tenya, and K. Machida, Phys. Rev. B **90**, 220502 (2014).
[23] E. Hassinger, P. Bourgeois-Hope, H. Taniguchi, S. René de Cotret, G. Grissonnanche, M. S. Anwar, Y. Maeno, N. Doiron-Leyraud, and L. Taillefer, Phys. Rev. X **7**, 011032 (2017).
[24] J. Jang, D. G. Ferguson, V. Vakaryuk, R. Budakian, S. B. Chung, P. M. Goldbart, and Y. Maeno, Science **331**, 186 (2011).
[25] S. B. Chung, H. Bluhm, and E.-A. Kim, Phys. Rev. Lett. **99**, 197002 (2007).
[26] V. Vakaryuk and A. J. Leggett, Phys. Rev. Lett. **103**, 057003 (2009).
[27] V. Vakaryuk and V. Vinokur, Phys. Rev. Lett. **107**, 037003 (2011).
[28] Y. Maeno, T. Ando, Y. Mori, E. Ohmichi, S. Ikeda, S. NishiZaki, and S. Nakatsuji, Phys. Rev. Lett. **81**, 3765 (1998).
[29] M. S. Anwar, T. Nakamura, S. Yonezawa, M. Yakabe, R. Ishiguro, H. Takayanagi, and Y. Maeno, Scientific Reports **3**, 2480 (2013).
[30] Y. Ying, N. Staley, Y. Xin, K. Sun, X. Cai, D. Fobes, T. Liu, Z. Mao, and Y. Liu, Nat. Commun. **4** (2013).
[31] S. Kittaka, H. Taniguchi, S. Yonezawa, H. Yaguchi, and Y. Maeno, Phys. Rev. B **81**, 180510 (2010).
[32] C. W. Hicks, D. O. Brodsky, E. A. Yelland, A. S. Gibbs, J. A. N. Bruin, M. E. Barber, S. D. Edkins, K. Nishimura, S. Yonezawa, Y. Maeno, and A. P. Mackenzie, Science **344**, 283 (2014).
[33] A. Steppke, L. Zhao, M. E. Barber, T. Scaffidi, F. Jerzembeck, H. Rosner, A. S. Gibbs, Y. Maeno, S. H. Simon, A. P. Mackenzie, and C. W. Hicks, Science **355** (2017).
[34] W. A. Little and R. D. Parks, Phys. Rev. Lett. **9**, 9 (1962).
[35] M. S. Anwar, R. Ishiguro, T. Nakamura, M. Yakabe, S. Yonezawa, H. Takayanagi, and Y. Maeno, Phys. Rev. B **95**, 224509 (2017).
[36] Y. Krockenberger, M. Uchida, K. Takahashi, M. Nakamura, M. Kawasaki, and Y. Tokura, Appl. Phys. Lett. **97**, 2502 (2010).
[37] J. Cao, D. Massarotti, M. E. Vickers, A. Kursumovic, A. D. Bernardo, J. W. A. Robinson, F. Tafuri, J. L. MacManus-Driscoll, and M. G. Blamire, Supercond. Sci. Technol. **29**, 095005 (2016).
[38] X. Cai, Y. A. Ying, N. E. Staley, Y. Xin, D. Fobes, T. J. Liu, Z. Q. Mao, and Y. Liu, Phys. Rev. B **87**, 081104 (2013).
[39] X. Cai, Y. Ying, B. Zakrzewski, D. Fobes, T. Liu, Z. Mao, and Y. Liu, arXiv:1507.00326.
[40] Z. Mao, Y. Maeno, and H. Fukazawa, Mater. Res. Bull. **35**, 1813 (2000).
[41] S. Yonezawa, T. Higuchi, Y. Sugimoto, C. Sow, and Y. Maeno, Rev. Sci. Instrum. **86**, 093903 (2015).
[42] See Supplemental Material [url] for the details of device fabrication process and additional results to support the conclusion of the main paper, which includes Refs. [48–52].

[43] G. R. Berdiyorov, S. H. Yu, Z. L. Xiao, F. M. Peeters, J. Hua, A. Imre, and W. K. Kwok, Phys. Rev. B **80**, 064511 (2009).

[44] V. V. Moshchalkov, L. Gielen, C. Strunk, R. Jonckheere, X. Qiu, C. V. Haesendonck, and Y. Bruynseraede, Nature **373**, 319 (1995).

[45] M. Morelle, D. S. Golubović, and V. V. Moshchalkov, Physical Review B **70**, 144528 (2004).

[46] V. V. Moshchalkov and J. Fritzsche, *Nanostructured Superconductors* (World Scientific Publishing, 2011), and the references there in.

[47] In-plane field value is also corrected using $H_y = H_y^{\mathrm{magnet}}\cos\theta - H_z^{\mathrm{magnet}}\sin\theta$. The mixed component $\mu_0 H_z^{\mathrm{magnet}}\sin\theta$ is only 0.078 mT even at the highest $H_z^{\mathrm{magnet}}$ value, which is comparable with the geomagnetic field (∼0.05 mT). Therefore, the in-plane magnetic field can be regarded as constant during the $H_z$ sweep.

[48] R. Loetzsch, A. Lübcke, I. Uschmann, E. Förster, V. Große, M. Thuerk, T. Koettig, F. Schmidl, and P. Seidel, Appl. Phys. Lett. **96**, 1901 (2010).

[49] O. Chmaissem, J. D. Jorgensen, H. Shaked, S. Ikeda, and Y. Maeno, Phys. Rev. B **57**, 5067 (1998).

[50] M. Lucht, M. Lerche, H.-C. Wille, Y. V. Shvyd'Ko, H. Rüter, E. Gerdau, and P. Becker, J. Appl. Cryst. **36**, 1075 (2003).

[51] D. Batchelder and R. Simmons, J. Chem. Phys. **41**, 2324 (1964).

[52] A. P. Mackenzie, R. K. W. Haselwimmer, A. W. Tyler, G. G. Lonzarich, Y. Mori, S. Nishizaki, and Y. Maeno, Phys. Rev. Lett. **80**, 161 (1998).

Supplemental Material for

# Little-Parks oscillations with half-quantum fluxoid features in $Sr_2RuO_4$ micro rings

Yuuki Yasui,[1] Kaveh Lahabi,[2] Muhammad Shahbaz Anwar,[1,3] Yuji Nakamura,[1]
Takahito Terashima,[1] Shingo Yonezawa,[1] Jan Aarts,[2] and Yoshiteru Maeno[1]

[1]*Department of Physics, Graduate School of Science, Kyoto University, Kyoto 606-8502, Japan*
[2]*Huygens-Kamerlingh Onnes Laboratory, Leiden University, P.O. Box 9504, 2300 RA Leiden, The Netherlands*
[3]*London Centre for Nanotechnology, University College London,*
*17-19 Gordon Street, London, WC1H 0AH, United Kingdom*

In this Supplemental Material, we describe the details of device fabrication process in Sec. I. Furthermore, we present additional results to support the conclusion of the main paper in Sec. II.

## I. PREPARATION OF THE MICRO RINGS

This section describes a detailed protocol of the fabrication of micro-ring devices using single crystalline $Sr_2RuO_4$. First, we crash a $Sr_2RuO_4$ crystal ($\sim 2 \times 2 \times 1$ mm$^3$) with tweezers to obtain small flakes, and select a thin flake with the size of about $30 \times 5 \times 1\ \mu$m$^3$ [Fig. S1(a)]. Second, as shown in Figs. S1(b), S1(c), and S2, we prepare electrodes using high-temperature-cure silver paint (Dupont, 6838). Third, we deposit a 100 nm-thick $SiO_2$ layer on top of the $Sr_2RuO_4$ crystal using the electron beam deposition technique to protect the crystals from the ion beam during the focused ion beam (FIB) process. $SiO_2$ deposition is performed under $\sim 1.8 \times 10^{-4}$ Pa with ~25 mA electron beam current, and the deposition rate is typically 0.3 nm/sec. Finally, as shown in Figs. S1(d) - (f) and Fig. S3, we mill the crystal using FIB to achieve a ring device. When we cut a large area, we first keep small regions not milled as shown in black in Figs. S3(d). This regions work as “walls” that prevent the redeposition to the ring. This wall is removed in the final part of the FIB process [Fig. S3(e)].

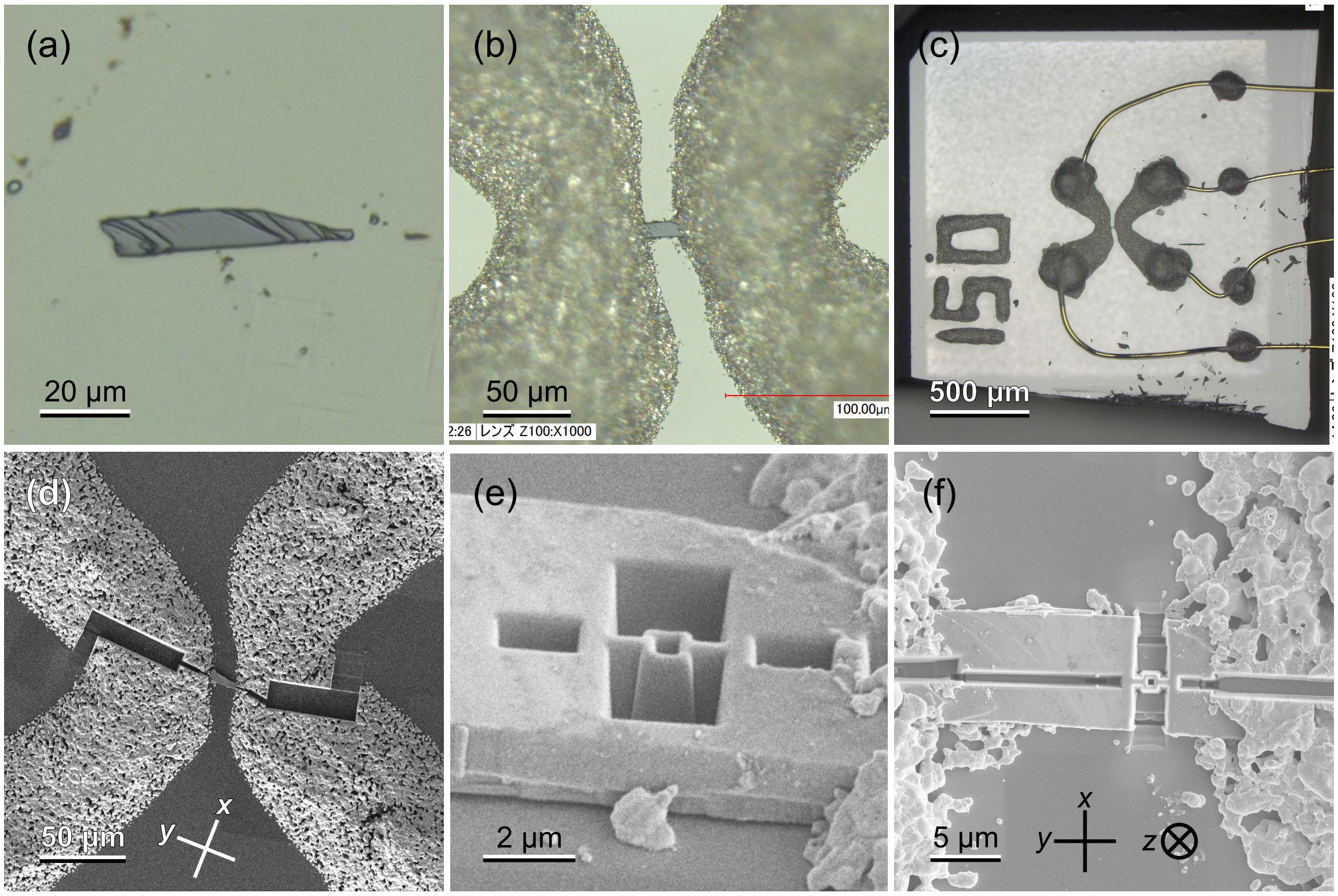

FIG. S1. Optical microscope images (a) - (c) and scanning electron microscope (SEM) images (d) - (f) of Sample B. (a) 50-$\mu$m long $Sr_2RuO_4$ single crystal. (b) Crystal connected to silver-paint electrodes. (c) Device after connecting gold wires prior to $SiO_2$ deposition. (d) Four-wire configuration after the silver parts are cut with FIB. (e) Device after the ring part is milled out. (f) Micro ring device after the completion of FIB process. A magnification of the central part is shown in Fig. 2(c) of the main paper.

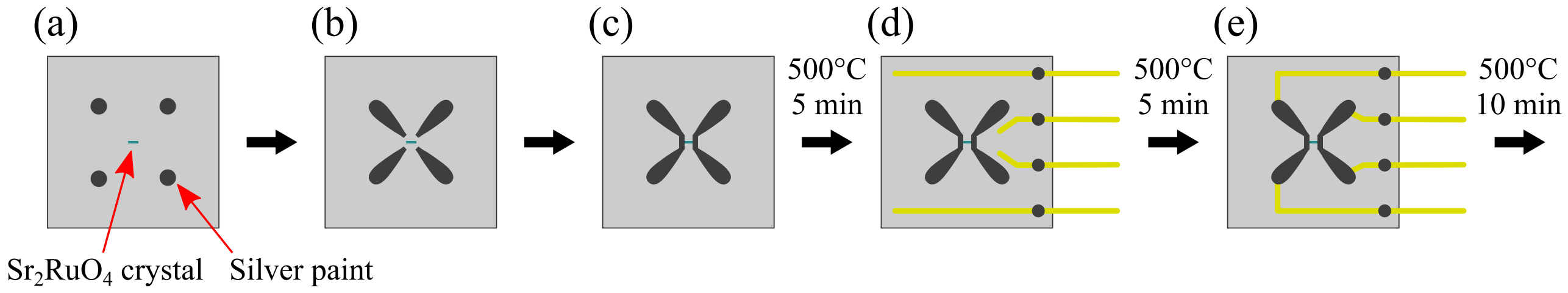

FIG. S2. Description of the process to prepare electrodes using high-temperature-cure silver paint. (a) $Sr_2RuO_4$ thin crystal placed at the center of a substrate, and four drops of silver paint also put on the substrate. (b) After the silver paint is pushed close to the crystal using a needle attached to a die bonder (West·Bond, 7200CR). (c) After the silver paint is connected to the crystal from both sides. In order to prevent the crystal from being sucked into the silver paint, it is important to connect both sides gradually and almost simultaneously. Then the silver paint is cured at 500°C for 5 min [Fig. S1(b)]. (d) With gold wires fixed on the substrate with additional silver paint. The silver paint is cured again. (e) After the gold wires connected to the silver electrodes using silver paint again. Finally, the silver paint is cured at 500°C for 10 min [Fig. S1(c)].

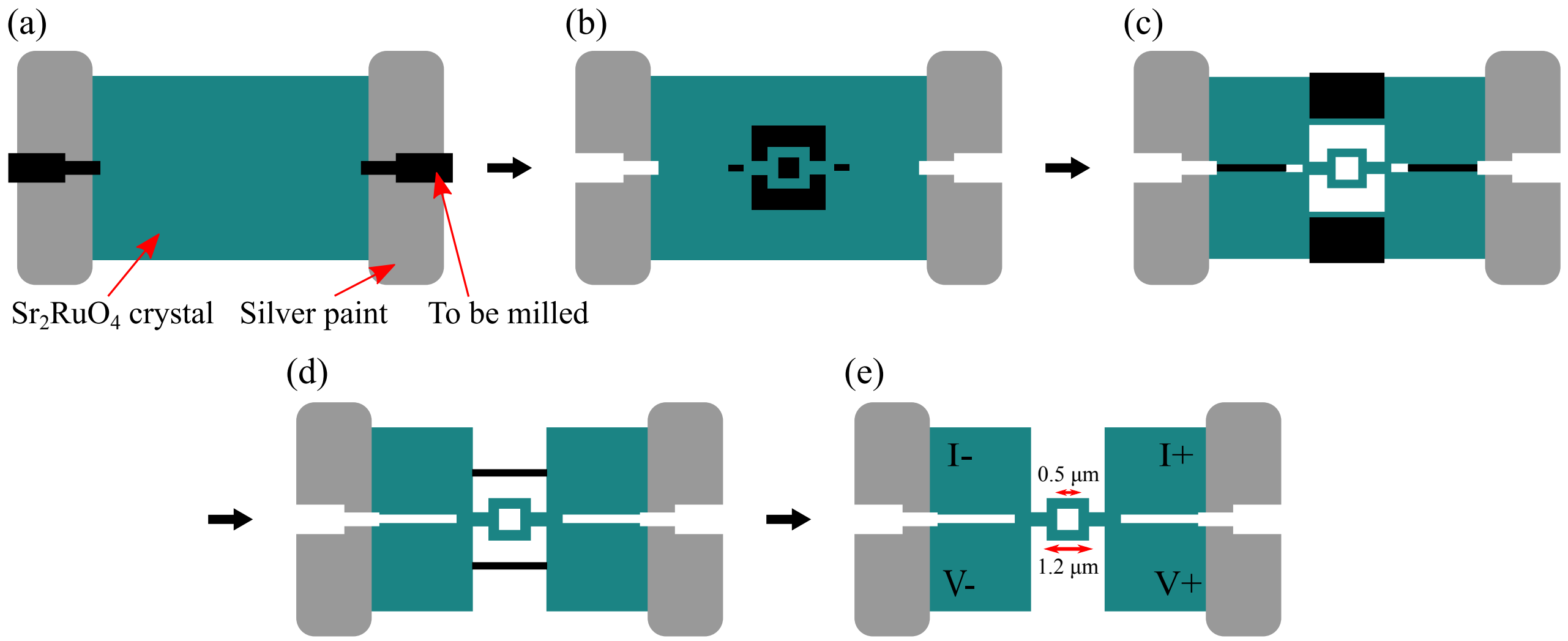

FIG. S3. Description of the process of milling out the ring with FIB. The black region is to be milled, and the white region represents the region that has been milled. (a) Before silver electrodes are cut with 16.4-nA, 30-kV beam to make four-wire configuration [Fig. S1(d)]. (b) Before the ring part is milled out with 20-pA, 30-kV beam as shown in Fig. S1(e). (c) Before the current- and voltage-terminals are separated. (d) Before the wall that worked to prevent the redeposition is removed. (e) Final configuration of the micro ring [Fig. S1(f)].

For this study, crystals of the batch numbers C89 and C391, which were grown by Z. Q. Mao and F. Hübler at Kyoto University, were used for Sample A (yy075) and Sample B (yy150), respectively. In earlier samples, including Sample A, we used sapphire ($\alpha$-$Al_2O_3$) substrates, but in most cases the rings broke during the cooling process from room temperature. For the later samples, including Sample B, $SrTiO_3$ substrates are selected because this oxide has similar thermal expansion (−0.19% along the $a$ axis from 300 K to 10 K [S1]) to that of $Sr_2RuO_4$ (−0.23% along the $a$ axis from 300 K to 15 K [S2]). For comparison, other substrates shrink less: $\alpha$-$Al_2O_3$: −0.06% along the $a$ axis from 300 K to 10 K [S3]; Si: −0.016% from 273 K to 6.4 K [S4]. This selection minimizes thermal damages or strains to the $Sr_2RuO_4$ micro rings during the cooling process from room temperature. Though $SrTiO_3$ shows cubic-to-tetragonal structural transition at 105 K [S1], we did not observe any anomaly in the $Sr_2RuO_4$ ring resistance due to this transition. An utmost care is taken during cooling process because sudden thermal contraction may break the rings. We typically spend two days to cool the samples from 300 K to 4 K by controlling the temperature with a heater.

## II. ADDITIONAL RESULTS

In this section, we present additional data supporting our scenario. In Subsection A, AC susceptibility of the $Sr_2RuO_4$ single crystal is shown to ensure the good quality of the crystal. In Subsection B, the origin of several superconducting resistance transition steps and the reason for the choice of the temperature for the magnetoresistance measurements are discussed. In Subsection C, detailed magnetoresistance and magnetovoltage related to Fig. 3 of the main paper is shown. In Subsection D, magnetoresistance and magnetovoltage data of Sample B, which are not included in the main paper, are presented for comparison.

### A. Superconducting Transition of a Crystal Before Ring Fabrication

The AC susceptibility of the $Sr_2RuO_4$ crystal C391, which was used for Sample B after crushed into small pieces, was measured with a compact susceptometer [S5] used with an adiabatic demagnetization refrigerator (ADR) option for Quantum Design PPMS. A sharp superconducting transition was observed at 1.50 K, which is the pure limit of the $T_c$ of $Sr_2RuO_4$ [S6]. However, the resistance onset $T_c$ of Sample B is substantially higher than 1.5 K [Fig. 2(e) and Fig. S6(a)]. This enhancement is most likely due to lattice strains caused by FIB process, considering that $T_c$ of pure $Sr_2RuO_4$ is known to increase with strains [S7, S8].

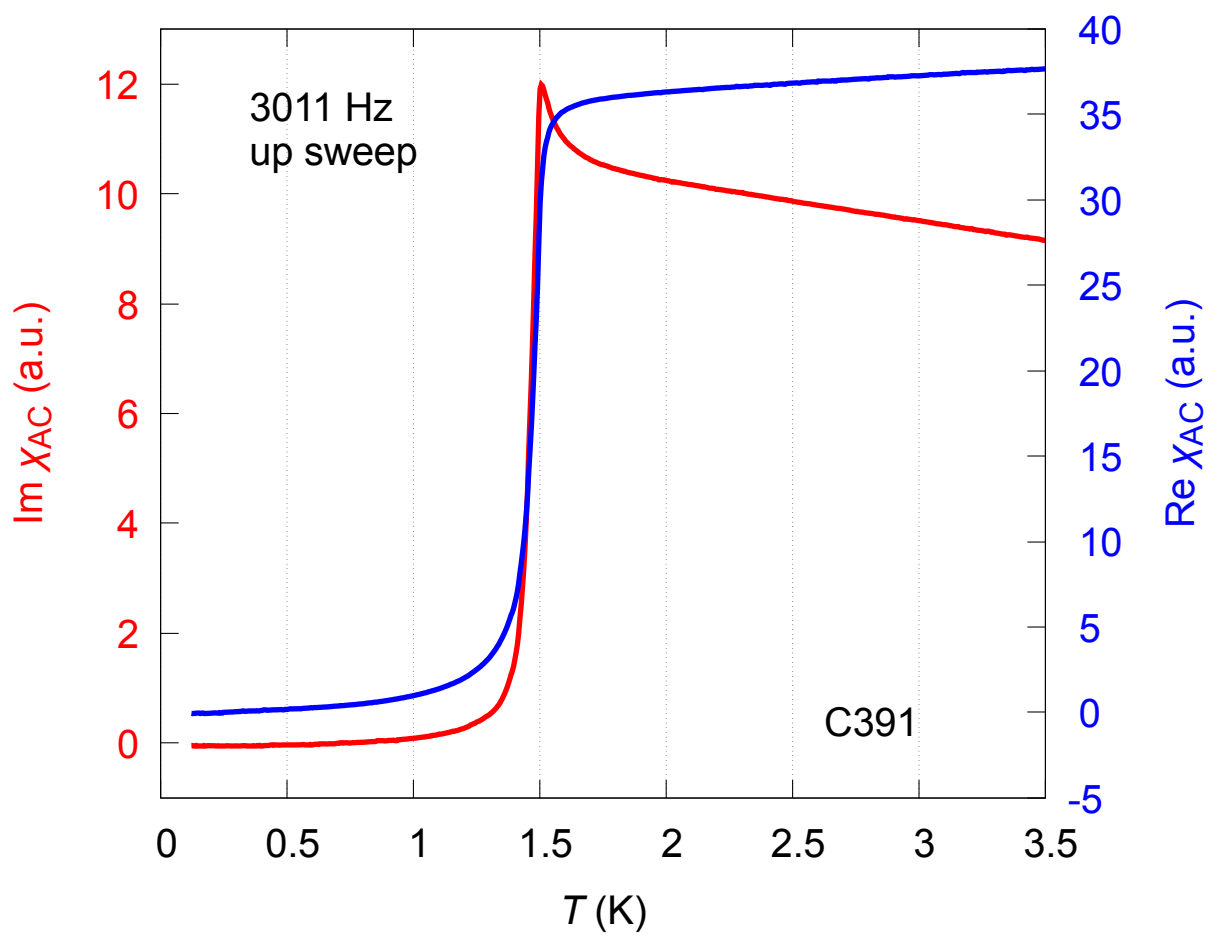

FIG. S4. AC susceptibility of the crystal C391. A sharp transition at $T_c$ = 1.50 K is observed.

### B. Temperature Stability during Magnetotransport Measurement

Although we use a $^3$He refrigerator (Oxford Instruments, model Heliox $2^{VL}$), the main measurements were performed at about 1.5 K. Since it is essential for the LP experiments to be performed under a high temperature stability, we condensed the liquid $^3$He in the $^3$He-pot without pumping to utilize its large thermal mass. In addition, we use a heater and thermometer on the sample stage to control the sample temperature precisely.

In order to reduce the noise in the magnetotransport measurements as well as to prevent the RF heating of the sample, we use an RC low-pass filter with the cutoff of 16 kHz for each electrical lead at the sample stage. We use a standard setup for the DC four-wire resistance measurements with a voltage source (Yokogawa, model 7651) in conjunction with the series resistance and a nano-voltmeter (Keithley, model 2182) capable of 10-nV resolution.

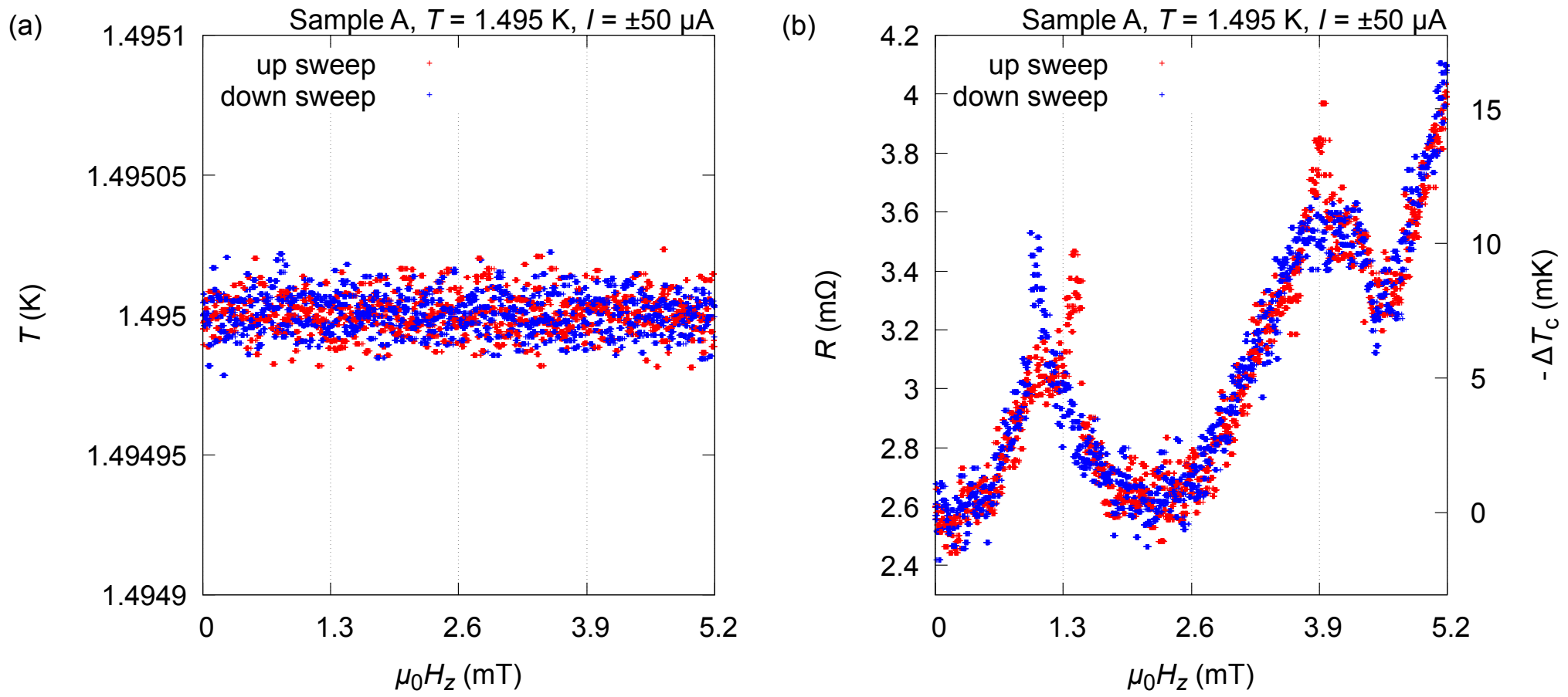

FIG. S5. (a) Temperature stability and (b) the corresponding resistance oscillations for Sample A during the magnetic-field sweep. The vertical axis on the right indicates the corresponding oscillations in $T_c$ derived from the $R(T)$ curve shown in Fig. 2(d): $-\Delta T_c = (dR/dT|_{T=1.495\ \mathrm{K}})^{-1}\,\Delta R$. Temperature stability in (a) is as small as 7.2 $\mu$K rms, much smaller than the oscillation amplitude in $T_c$ of about 10 mK. Such oscillation amplitude in $T_c$ is quantitatively consistent with that estimated from Eq. (1). The overall temperature stability is maintained within about 100 $\mu$K rms.

## C. Modeling for HQF

For the dashed curves in Fig. 3(d), we use the following model based on Eq. (1) in the main paper.

$$V_{\mathrm{IQF}}(H_z; H_y) = aH_z^2 + b\left(n - \frac{H_z}{\Delta H}\right)^2 + c_{H_y}, \tag{S1}$$

$$V_{\mathrm{HQF}}(H_z; H_y) = aH_z^2 + B\left(n - \frac{1}{2} - \frac{H_z}{\Delta H}\right)^2 + c_{H_y} + d_{H_y}, \tag{S2}$$

$$V(H_z; H_y) = \min\left[V_{\mathrm{IQF}}(H_z; H_y), V_{\mathrm{HQF}}(H_z; H_y)\right], \tag{S3}$$

where $n$ is an integer, and $\Delta H$ is the period of the oscillations. $V_{\mathrm{IQF}}, V_{\mathrm{HQF}}$ model the voltage corresponding to IQF and HQF states, respectively. Since the state with the lowest energy would correspond to the lowest voltage, we use Eq. (S3) to evaluate it between IQF and HQF states. The parameters $a, b, B, \Delta H$ are common for all $H_y$, while $c_{H_y}$ and $d_{H_y}$ are decided for each $H_y$ value.

## D. Magnetotransport of Sample B at Different Temperatures

As shown in Fig. S6(a), the temperature dependence of the resistance $R(T)$ of Sample B exhibits three transition steps in the temperature dependence of the resistance $R(T)$; the onset of each transition (5.5, 11, and 24 mΩ) is indicated with broken horizontal lines. Interestingly, these resistance values at the onset $T_c$'s do not depend on the magnitude of the measurement currents. To clarify the origin of such transition steps, we investigated the magnetoresistance at various temperatures [Fig. S6(b)]. The magnetoresistance behavior can be categorized into three groups with the boundary resistance $R$ values common to the temperature dependence; monotonic increase for $R < 5.5$ mΩ (the vertical green arrow), complicated behavior with noisy features for 5.5 mΩ $< R <$ 11 mΩ (the blue arrow), and periodic oscillations for 11 mΩ $< R <$ 24 mΩ (the red arrow). Such changes in the behavior occurs when the resistance reaches the boundary values with increasing either the temperature or the magnetic field.

Such correspondence can be naturally explained if each transition step originates from independent superconducting transitions. Firstly, the magnetoresistance oscillation period at high temperatures agrees well with the magnetic flux quantum $\Phi_0 = h/2e$ for 0.54 $\mu m^2$ area, which corresponds to the ring geometry. This fact indicates that the first transition (the red region) originates from the superconductivity of the ring part. We comment here that this ring part is expected to be most affected by the ion beam, and the 3K-phase is most likely induced. The regions corresponding to the green and blue transitions are not easy to assign. Nevertheless, because the contact part is probably less exposed to the ion beam, we assign the green transition to the superconductivity in the contact part. Finally, we attribute the blue transition to the neck part.

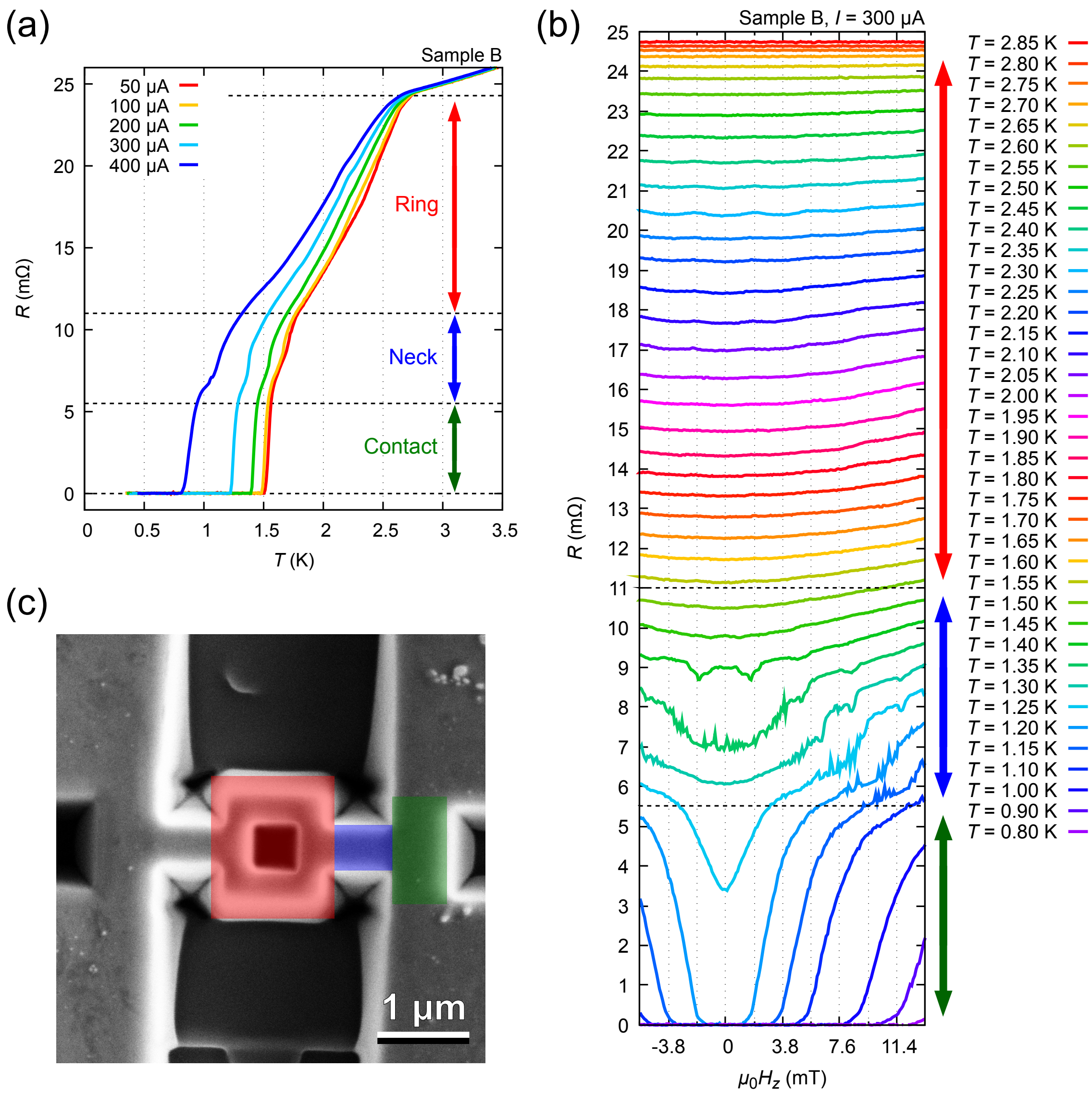

FIG. S6. (a) Temperature dependence of the resistance of Sample B with various measurement current. Three transition steps are observed, whose resistance values at the onset $T_c$'s are indicated with dashed lines. (b) Magnetoresistance of Sample B at various temperatures. The dashed lines represent the boundaries of the different behaviors. (c) The red, blue, and green transitions correspond to those of the ring-, the neck-, and the contact-part, respectively.

Based on such assignment, we chose 2.3 K for the magnetotransport measurement with and without in-plane field. This is because the transition originates from the superconductivity of the ring part and 2.3 K is well separated from the other transitions.

### E. Detailed Data of Sample A

Here, we present the magnetoresistance and magnetovoltage of Sample A with different in-plane field directions $x$ and $y$. These measurements were performed in the same measurement conditions as the measurement explained in Fig. 3 of the main paper. The magnetoresistance and magnetovoltage of Sample A with positive and negative $H_y$ are presented in Figs. S7 and S8, respectively. Similarly, those for in-plane field along the $x$ direction are shown in Fig. S9 (positive $H_x$) and Fig. S10 (negative $H_x$). In these figures, the panels (a) show the ring resistance, the panels (b) show the sample voltage under positive current $V_+ = V(+I)$, and the panels (c) show the sample voltage under negative current $V_- = V(-I)$. Notice that $R$ is evaluated as $R = (V_+ - V_-)/2I$. In addition, we point out that the expected symmetry $V_\pm(\boldsymbol{H}) \simeq -V_\mp(-\boldsymbol{H})$ is satisfied both for in-plane fields along $x$ and along $y$ directions.

As explained in the main paper, the splitting of the oscillation peaks, i.e. the HQF feature, was observed under $H_y$ as indicated with arrows in Figs. S7 and S8 in panels (b) and (c). Such HQF feature is observed in 2 data sets (see Fig. S7(b) red curves and Fig. S8(c) green curves) out of 8 possible combinations ($2^3$ combinations for the signs of the directions of $H_y$ and $I$, as well as the sweep directions of $H_z$). In contrast, the HQF features are less clear for $H_x$ than for $H_y$. Such difference may be attributed to the breaking of the in-plane rotational symmetry due to the measurement current and the sample geometry, namely the Lorentz force. Because the in-plane field and the measurement current are perpendicular for $H_x$, the Lorentz force affect the magnetoresistance more strongly than for $H_y$.

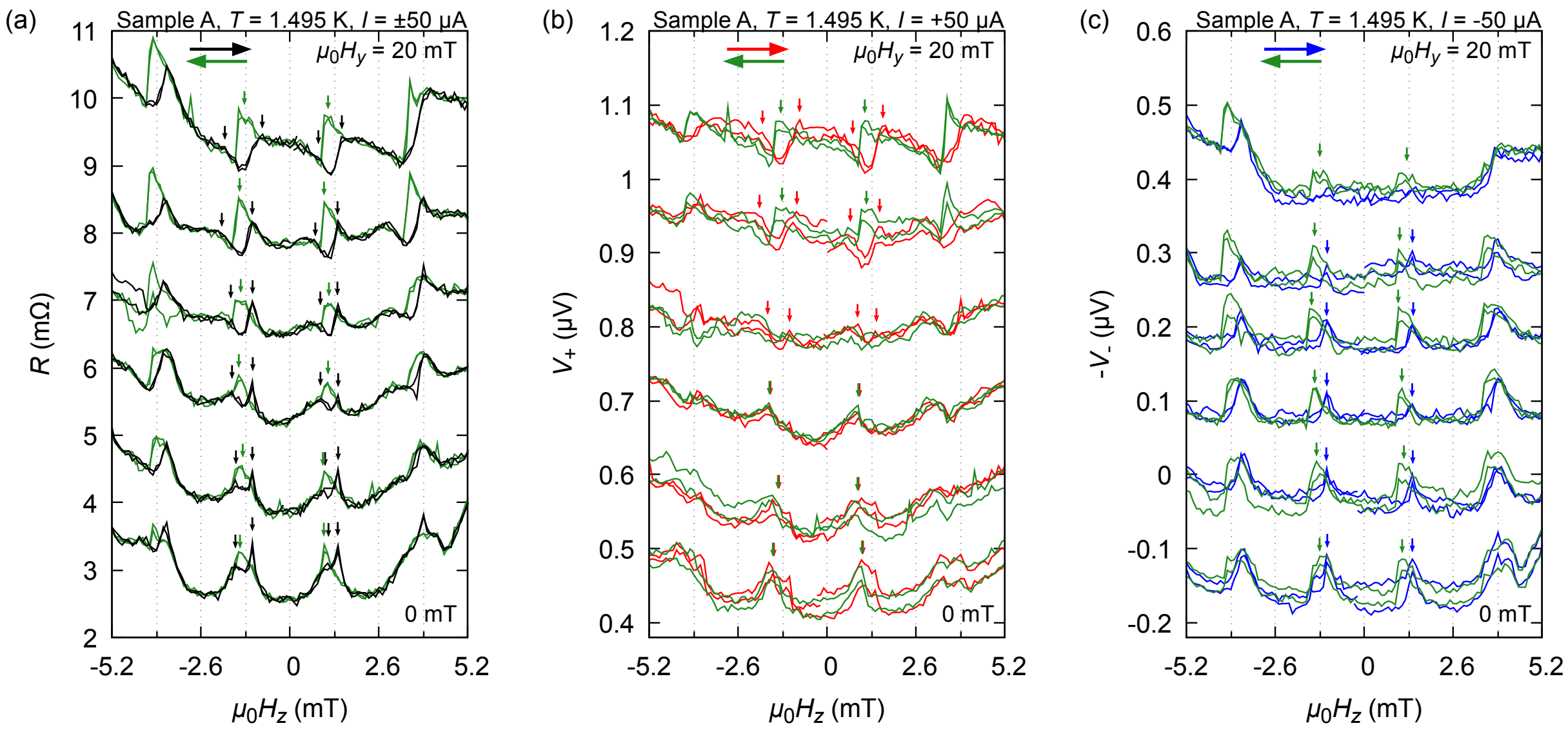

FIG. S7. Magnetoresistance and magnetovoltage of Sample A with positive $H_y$. The $y$ direction corresponds to that shown in Fig. S1(f). (a) Magnetoresistance $R(H_z, H_y)$ with every 4-mT $H_y$ steps. Each set of the curves has a 1-mΩ offset. (b) Magnetovoltage when current is applied to the positive direction $V_+(H_z, H_y)$. In the field up-sweep (red), also presented in Fig. 3(d) of the main paper, a qualitative change from the magnetovoltage peak to dip occurs between 8 and 12 mT. In the field down-sweep (green), a corresponding change occurs but from the peak to a sudden magnetovoltage drop at a field corresponding to the lower field side of the dip. (c) Magnetovoltage when current is applied to the negative direction $V_-(H_z, H_y)$. In this set of current-field orientations, a qualitative change with respect to the magnitude of $H_y$ does not occur. Nevertheless, a hysteresis between field up- and down-sweeps is observed. Each set of the curves has a 0.1-$\mu$V offset.

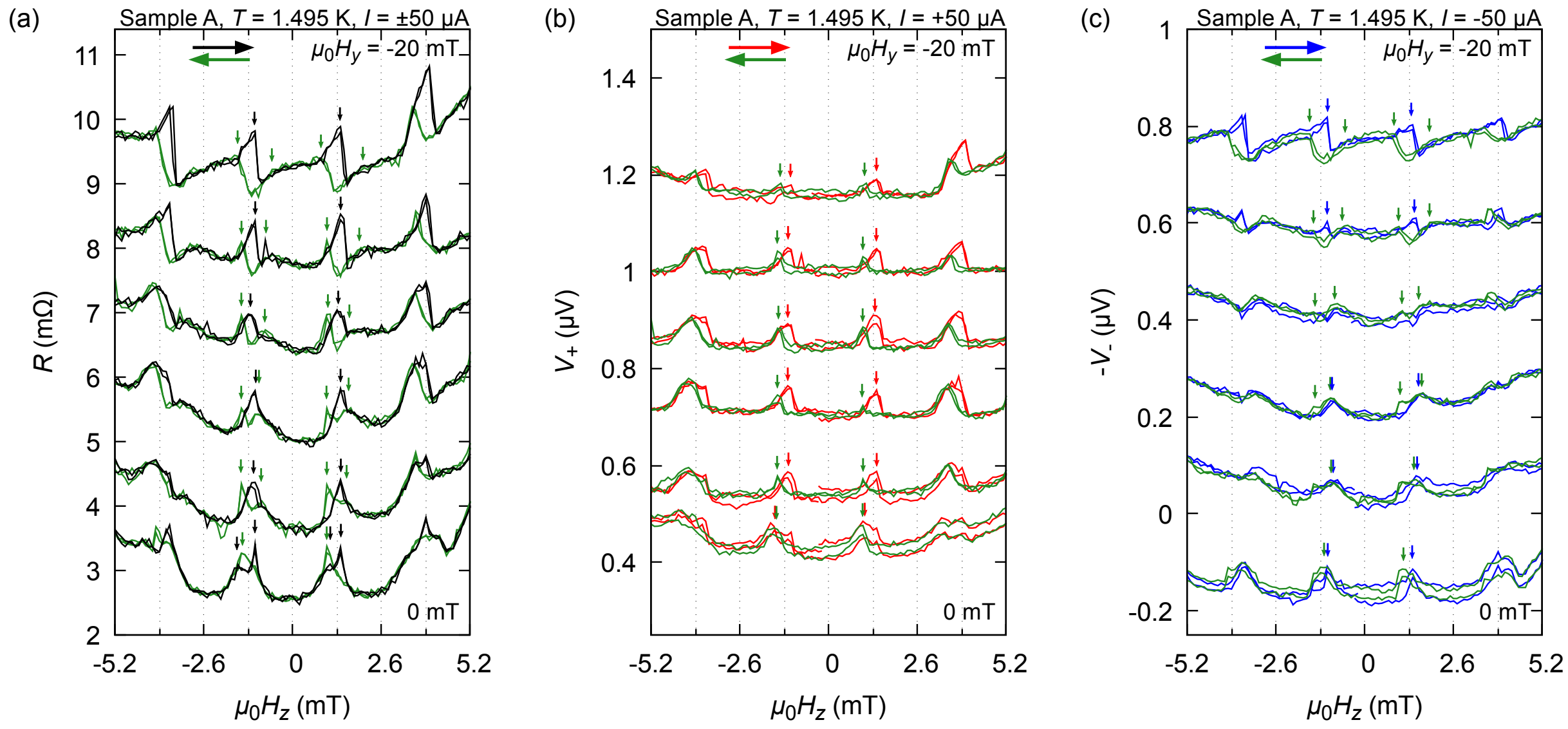

FIG. S8. Magnetoresistance and magnetovoltage of Sample A with negative $H_y$. The $y$ direction corresponds to that shown in Fig. S1(f). (a) Magnetoresistance $R(H_z, H_y)$ with every 4-mT $H_y$ steps. Each set of the curves has a 1-mΩ offset. (b) Magnetovoltage when current is applied to the positive direction $V_+(H_z, H_y)$. Similarly to Fig. S7(c), a qualitative change with respect to the magnitude of $H_y$ does not occur, and a hysteresis between field up- and down-sweeps is observed. (c) Magnetovoltage when current is applied to the negative direction $V_-(H_z, H_y)$. Similarly to Fig. S7(b), in the field down-sweep (green) a qualitative change from the magnetovoltage peak to dip occurs between 8 and 12 mT. In the field up-sweep (blue), a corresponding change occurs but from the peak to a sudden magnetovoltage drop at a field corresponding to the lower field side of the dip. Each set of the curves has a 0.15-$\mu$V offset.

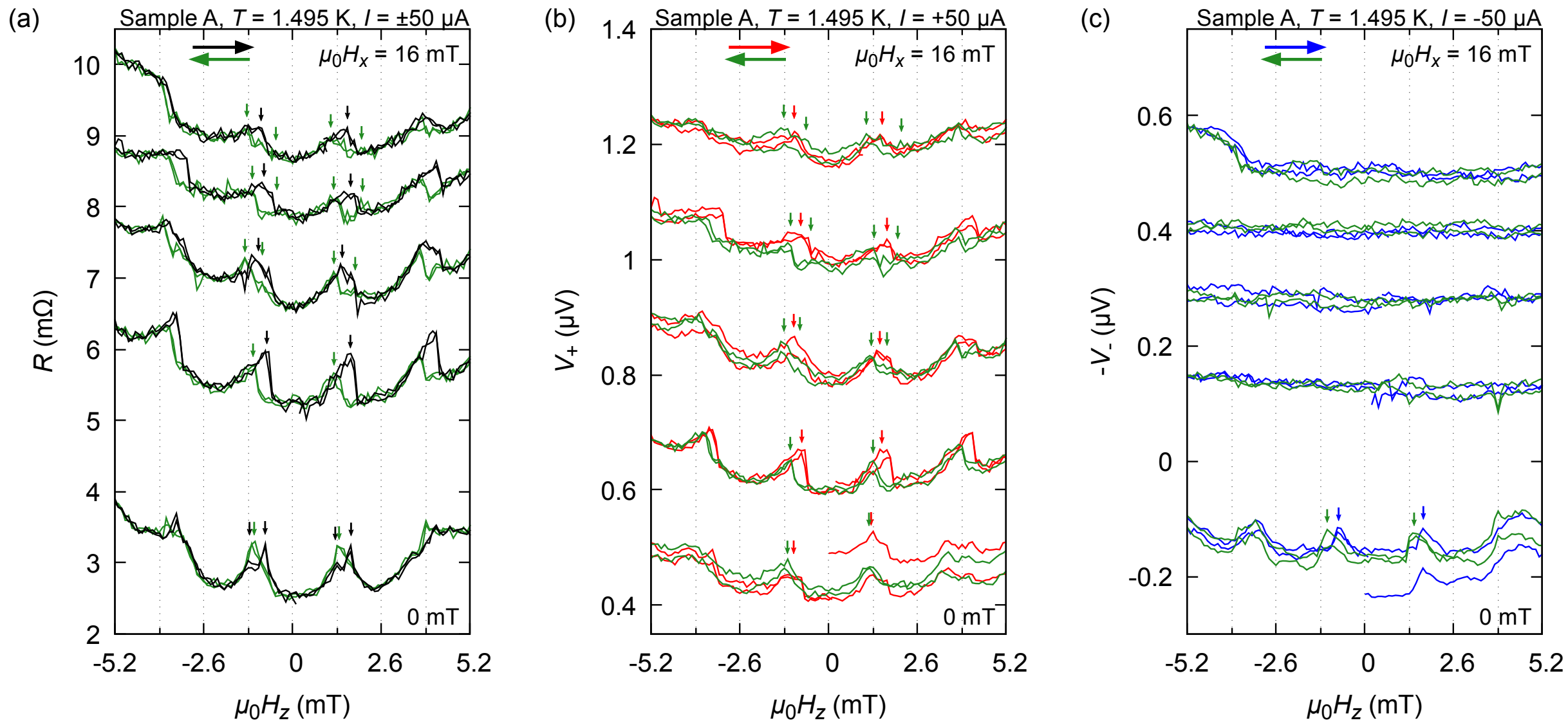

FIG. S9. Magnetoresistance and magnetovoltage of Sample A with positive $H_x$. The $x$ direction corresponds to that shown in Fig. S1(f). (a) Magnetoresistance $R(H_z, H_x)$ with every 4-mT $H_x$ steps. Each set of the curves has a 0-mΩ offset. (b) Magnetovoltage when current is applied to the positive direction $V_+(H_z, H_x)$. A very similar behavior to Fig. S6(b) with $H_x$ occurs. The qualitative change from the peak to dip occurs between 4 and 8 mT. (c) Magnetovoltage when current is applied to the negative direction $V_-(H_z, H_x)$. The oscillations disappear immediately upon application of $H_x$. This means the magnetoresistance oscillations originate entirely from magnetovoltage $V_+$. Each set the of the curves has a 0.1-$\mu$V offset.

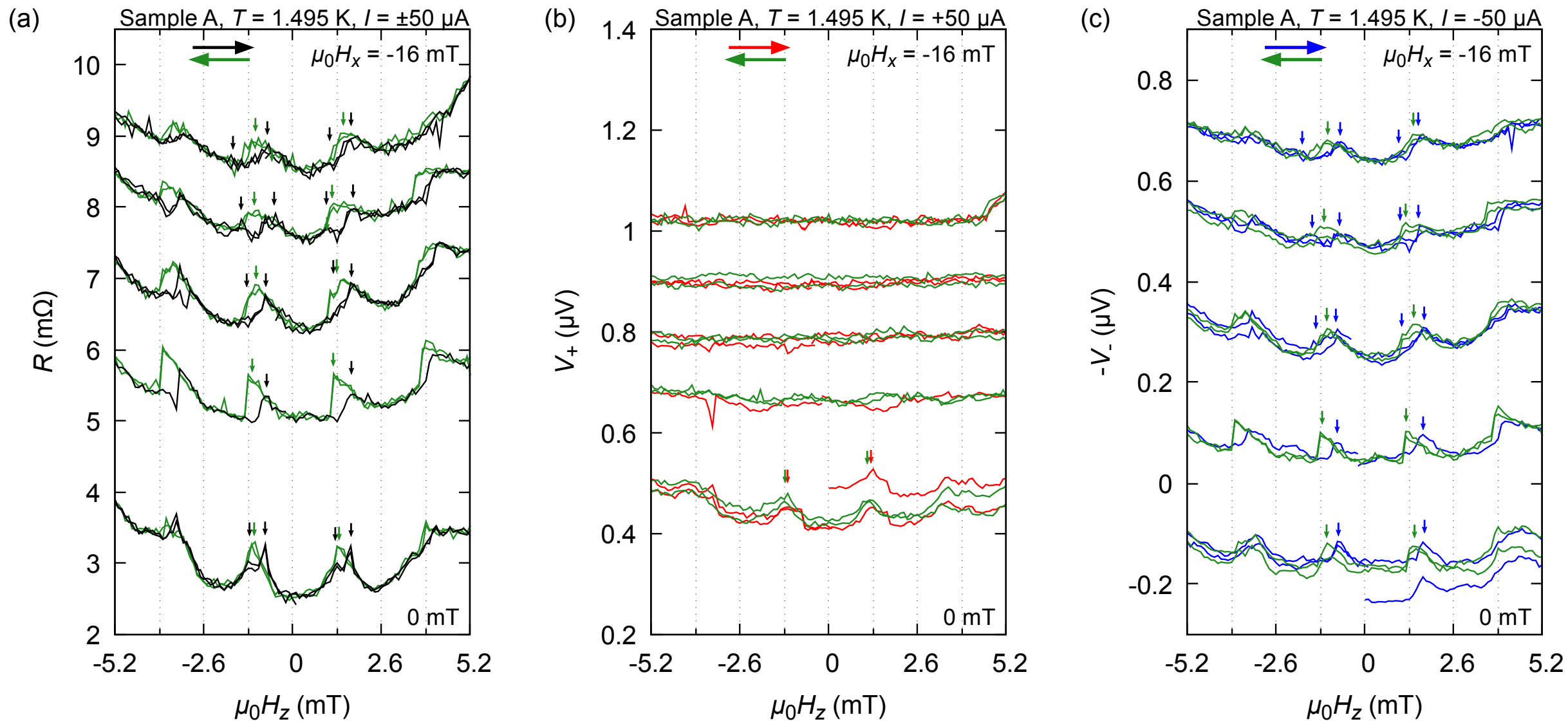

FIG. S10. Magnetoresistance and magnetovoltage of Sample A with negative $H_x$. The $x$ direction corresponds to that shown in Fig. S1(f). (a) Magnetoresistance $R(H_z, H_x)$ with every 4-mT $H_x$ steps. Each set of curves has a 0-mΩ offset. (b) Magnetovoltage when current is applied to the positive direction $V_+(H_z, H_x)$. Similarly to Fig. S9(c), the oscillations disappear immediately upon application of $H_x$, meaning the magnetoresistance oscillations originate entirely from magnetovoltage $V_-$. (c) Magnetovoltage when current is applied to the negative direction $V_-(H_z, H_x)$. Each set of the curves has a 0.1-$\mu$V offset.

### F. Magnetovoltage Peak Splitting in Sample B

In the main paper, we present magnetoresistance and magnetovoltage of Sample B without in-plane field [Fig. 3(b)]. Here, we present the magnetotransport with in-plane fields (Figs. S11 for positive $H_y$ and S12 for negative $H_y$). The splitting of the oscillation peaks is clearly seen as indicated with the arrows: with increasing in-plane field, new resistance minima appear at the peak position and the splitting width increases. These features are qualitatively consistent with the expectation that the HQF states are stabilized with the in-plane field. Nevertheless, some of the observed features cannot be explained within the simple HQF scenario. First, the in-plane field value where the splitting begins is substantially larger than those of Sample A and of Jang's report [S9], and is even larger than the in-plane $H_{c1}$ of thin crystals ($\simeq$25 mT at 0.5 K [S9]). Second, additional peaks appear near $\mu_0 H_z = 0$ mT. Third, the resistance becomes unstable at around $\mu_0 H_y \sim 154$ and $-152$ mT. Since the applied in-plane field is larger than the in-plane $H_{c1}$, in-plane vortices must affect these features. It is not clearly known how the HQF features are affected by in-plane vortices. Note that the applied in-plane field is still smaller than the in-plane $H_{c2}$ because the resistance is approximately 20.8 mΩ, which is lower than the normal state resistance 24 mΩ.

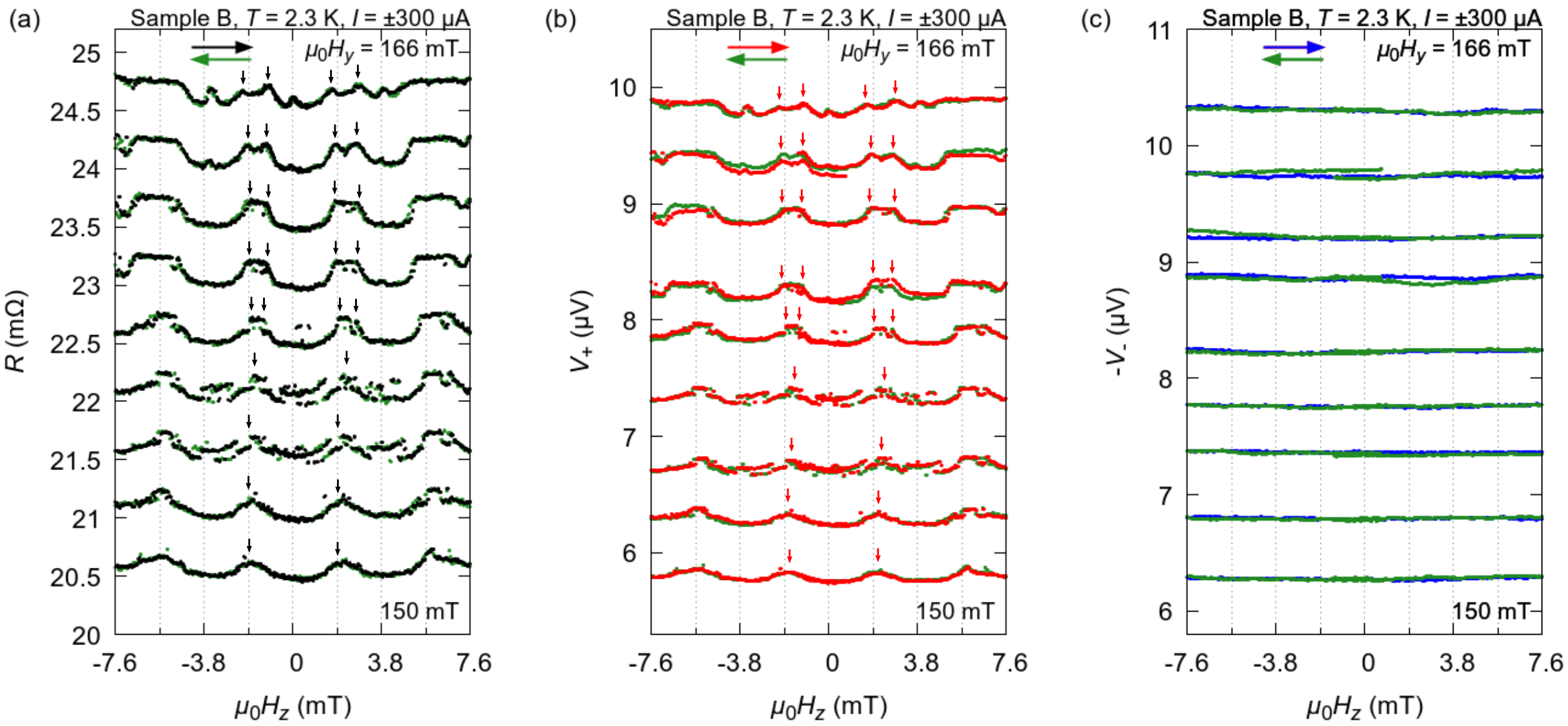

FIG. S11. Magnetoresistance and magnetovoltage of Sample B with positive $H_y$. (a) Magnetoresistance $R(H_z, H_y)$ with every 2-mT $H_y$ steps. Each set of the curves has a 0.5-mΩ offset. (b) Magnetovoltage when current is applied to the positive direction $V_+(H_z, H_y)$. The peak splitting at $\Phi_0/2$ is clearly observed as expected for HQF states. (c) Magnetovoltage when current is applied to the negative direction $V_-(H_z, H_x)$. $V_-$ hardly changes as a function of $H_z$. This means the magnetoresistance oscillations originate entirely from $V_+$. Each set of the curves has a 0.5-$\mu$V offset.

[S1] R. Loetzsch, A. Lübcke, I. Uschmann, E. Förster, V. Große, M. Thuerk, T. Koettig, F. Schmidl, and P. Seidel, Appl. Phys. Lett. **96**, 1901 (2010).

[S2] O. Chmaissem, J. D. Jorgensen, H. Shaked, S. Ikeda, and Y. Maeno, Phys. Rev. B **57**, 5067 (1998).

[S3] M. Lucht, M. Lerche, H.-C. Wille, Y. V. Shvyd'Ko, H. Rüter, E. Gerdau, and P. Becker, Journal of applied crystallography **36**, 1075 (2003).

[S4] D. Batchelder and R. Simmons, The Journal of Chemical Physics **41**, 2324 (1964).

[S5] S. Yonezawa, T. Higuchi, Y. Sugimoto, C. Sow, and Y. Maeno, Rev. Sci. Instrum. **86**, 093903 (2015).

[S6] A. P. Mackenzie, R. K. W. Haselwimmer, A. W. Tyler, G. G. Lonzarich, Y. Mori, S. Nishizaki, and Y. Maeno, Phys. Rev. Lett. **80**, 161 (1998).

[S7] S. Kittaka, H. Taniguchi, S. Yonezawa, H. Yaguchi, and Y. Maeno, Phys. Rev. B **81**, 180510 (2010).

[S8] C. W. Hicks, D. O. Brodsky, E. A. Yelland, A. S. Gibbs, J. A. N. Bruin, M. E. Barber, S. D. Edkins, K. Nishimura, S. Yonezawa, Y. Maeno, and A. P. Mackenzie, Science **344**, 283 (2014).

[S9] J. Jang, D. G. Ferguson, V. Vakaryuk, R. Budakian, S. B. Chung, P. M. Goldbart, and Y. Maeno, Science **331**, 186 (2011).

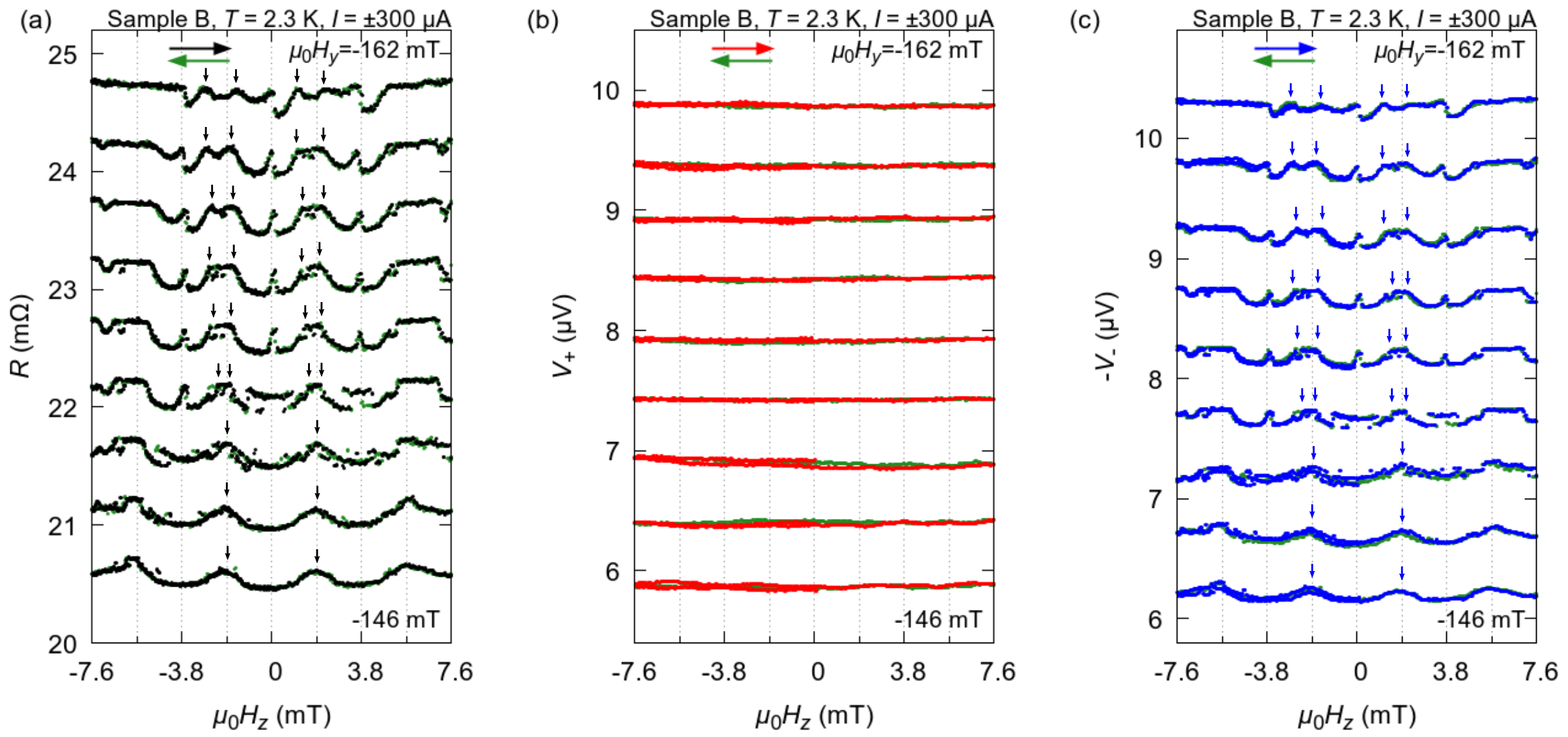

FIG. S12. Magnetoresistance and magnetovoltage of Sample B with negative $H_y$. (a) Magnetoresistance $R(H_z, H_y)$ with every 2-mT $H_y$ steps. Each set of the curves has a 0.5-mΩ offset. (b) Magnetovoltage when current is applied to the positive direction $V_+(H_z, H_y)$. Similarly to Fig. S11(c), $V_+$ hardly changes as a function of $H_z$, meaning the magnetoresistance oscillations originate entirely from $V_-$. (c) Magnetovoltage when current is applied to the negative direction $V_-(H_z, H_x)$. Similarly to Fig. S11(b), the peak splitting at $\Phi_0/2$ is clearly observed as expected for HQF states. Each set of the curves has a 0.5-$\mu$V offset.